\documentclass[twocolumn,superscriptaddress,showpacs,nofootinbib,notitlepage,preprintnumbers,secnumarabic,amssymb, floatfix, nobibnotes, aps, prd]{revtex4-2}
\usepackage[utf8]{inputenc}
\usepackage{graphicx}
\usepackage{latexsym,amsmath,amssymb,amsthm,lmodern,float,url}
\usepackage{natbib}
\usepackage{color}
\usepackage{microtype}
\usepackage{import}
\usepackage{bbold}
\usepackage[plain]{fancyref}
\usepackage{varioref}
\usepackage{subfigure}
\usepackage{slashed}
\usepackage{multirow}
\usepackage{tikz}
\usetikzlibrary{shapes}
\usetikzlibrary{positioning}
\usepackage[normalem]{ulem}
\usepackage{bbm}
\usepackage{qcircuit}
\usepackage[colorlinks=true,backref=false, linktocpage=true,
citecolor=blue,urlcolor=blue,linkcolor=blue,pdfpagemode=UseOutlines]{hyperref}
\hypersetup{%
  bookmarksnumbered=true,
  pdftitle = {},
  pdfsubject = {},
  pdfauthor = {},
  pdfkeywords = {}
}

\newcommand{\ket}[1]{\left| #1 \right>} 
\newcommand{\braket}[2]{\left< #1 \vphantom{#2} \right|
 \left. #2 \vphantom{#1} \right>} 
\newcommand{\mbraket}[3]{\left< #1 \vphantom{#2#3} \right|
 #2 \left| #3 \vphantom{#1#2} \right>} 

\newcommand{\jakarta}{\texttt{ibmq\_jakarta}}
\newcommand{\nairobi}{\texttt{ibm\_nairobi}}
\newcommand{\manila}{\texttt{ibmq\_manila}}
\newcommand{\quito}{\texttt{ibmq\_quito}}
\begin{document}
\preprint{FERMILAB-PUB-23-171-SQMS-T}
\title{Simulating $\mathbbm{Z}_2$ lattice gauge theory on a quantum computer}

\author{Clement Charles}
\affiliation{Department of Physics, The University of the West Indies, St. Augustine Campus, Trinidad \& Tobago}
\affiliation{Physics Division, Lawrence Berkeley National Laboratory, Berkeley, CA 94720, USA}

\author{Erik J. Gustafson}
\affiliation{Fermi National Accelerator Laboratory, Batavia,  Illinois, 60510, USA}
\affiliation{Quantum Artificial Intelligence Laboratory (QuAIL),
NASA Ames Research Center, Moffett Field, CA, 94035, USA}
\affiliation{USRA Research Institute for Advanced Computer Science (RIACS), Mountain View, CA, 94043, USA}

\author{Elizabeth Hardt}
\affiliation{Department of Physics, University of Illinois at Chicago, Chicago, Illinois 60607, USA}
\affiliation{Advanced Photon Source, Argonne National Laboratory, Argonne, Illinois 60439, USA}

\author{Florian Herren}
\affiliation{Fermi National Accelerator Laboratory, Batavia,  Illinois, 60510, USA}

\author{Norman Hogan}
\affiliation{Department of Physics, North Carolina State University, Raleigh, North Carolina 27695, USA}

\author{Henry Lamm}
\affiliation{Fermi National Accelerator Laboratory, Batavia,  Illinois, 60510, USA}

\author{Sara Starecheski}
\affiliation{Department of Physics, Sarah Lawrence College, Bronxville, NY 10708, USA}
\affiliation{Department of Physics,
University of Illinois at Urbana-Champaign, Urbana, IL 61801, USA}

\author{Ruth S. Van de Water}
\affiliation{Fermi National Accelerator Laboratory, Batavia,  Illinois, 60510, USA}

\author{Michael L. Wagman}
\affiliation{Fermi National Accelerator Laboratory, Batavia,  Illinois, 60510, USA}

\date{\today}
\begin{abstract}

The utility of quantum computers for simulating lattice gauge theories is currently limited by the noisiness of the physical hardware. Various quantum error mitigation strategies exist to reduce the statistical and systematic uncertainties in quantum simulations via improved algorithms and analysis strategies.
We perform quantum simulations of $1+1d$ $\mathbbm{Z}_2$ gauge theory with matter to study the efficacy and interplay of different error mitigation methods: readout error mitigation, randomized compiling, rescaling, and dynamical decoupling.
We compute Minkowski correlation functions in this confining gauge theory and extract the mass of the lightest spin-1 state from fits to their time dependence.
Quantum error mitigation extends the range of times over which our correlation function calculations are accurate by a factor of six and is therefore essential for obtaining reliable masses.
\end{abstract}

\maketitle

\section{Introduction}
Nucleons and other hadrons are ubiquitous in particle- and nuclear-physics experiments.
Predicting their properties theoretically requires a nonperturbative description of bound states in quantum chromodynamics (QCD).
Nonperturbative QCD uncertainties are now the dominant theoretical uncertainties in a wide range of new physics searches in areas ranging from quark and lepton flavor physics to neutron star physics to neutrino physics~\cite{Aoyama:2020ynm,Ruso:2022qes,Cirigliano:2022oqy,Tews:2022yfb}.
Lattice QCD has become a highly successful tool that enables nonperturbative calculations of QCD observables such as hadron masses and matrix elements with systematically improvable uncertainties that are essential to the success of many ongoing and future experiments~\cite{Drischler:2019xuo,Davoudi:2020ngi,Bulava:2022ovd,USQCD:2022mmc,Davoudi:2022bnl}.
However, lattice QCD calculations on classical computers exclusively rely on Monte Carlo algorithms to compute the high-dimensional path integrals required for precise calculations of hadron properties.
Monte Carlo algorithms are very efficient for path integrals whose integrands are positive definite, but the presence of sign problems -- integrands with sign or complex-phase fluctuations -- leads to computational resource requirements that grow exponentially with system size~\cite{Gattringer:2016kco}.
Many systems of interest cannot presently be studied with lattice QCD using classical computers because of challenges arising from sign problems, including all systems involving nonequilibrium real-time evolution.
Although there has been progress in developing novel classical lattice gauge theory (LGT) algorithms~\cite{Alexandru:2016gsd,Hoshina:2020gdy,Kanwar:2021tkd,Lawrence:2021izu}, simulations of the dynamics of four-dimensional LGTs relevant for studying hadronization of quarks and gluons, heavy-ion collisions, and phase transitions in the early universe remain unfeasible.

Quantum computers offer the possibility to simulate real-time dynamics of LGTs without sign problems by using Hamiltonian time evolution of quantum states instead of Monte-Carlo evaluations of path integrals~\cite{Feynman:1981tf}.  
As with the evolution of LGT on classical computers, however, realizing this potential will require a sustained effort over many years to understand and address theoretical issues, develop and test efficient algorithms, and identify and control systematic uncertainties.
Therefore, even though large-scale fault-tolerant quantum computers capable of realistic four-dimensional lattice gauge theory (LGT) simulations will not be realized for many years, it is timely to begin exploring the possibilities and tackling the challenges associated with quantum simulations of LGTs. 

Extracting accurate physical results from noisy intermediate-scale quantum (NISQ) hardware requires quantifying and reducing systematic uncertainties.
Substantial effort has been dedicated to developing and testing strategies for quantum error mitigation (QEM) in quantum simulations for chemistry, physics, and mathematics~\cite{Temme:2016vkz,10.1145/3386162,Ferracin:2022qqv,PhysRevA.99.032329,Alexandrou:2021ynh,Kim:2021gvc,Berg:2022ugn,Cai:2022rnq,Saki:2023rum,Ezzell2022,Kandala2017,Dumitrescu:2018njn,PhysRevX.8.011021,PhysRevResearch.2.043140,https://doi.org/10.48550/arxiv.2010.08520,Arute:2020uxm,TILLY20221,https://doi.org/10.48550/arxiv.2103.09846,Bennewitz:2021jqi,Chen:2022zxy,OBrien:2022uip,PhysRevC.106.034325,Gustafson:2023lhl}.
In LGT calculations, the use~\cite{Klco:2018kyo,Klco:2019evd,Ciavarella:2020vqm,Alam:2021uuq,Ciavarella:2021nmj,Ciavarella:2021lel,Atas:2022dqm,Carena:2022kpg,Parks:2022kdb,Gustafson:2022xdt} and systematic study~\cite{PhysRevD.106.094502,Zhao:2022dma,Stryker:2018efp,Gustafson:2019vsd,Halimeh:2019svu,Lamm:2020jwv,Tran:2020azk,Kasper:2020owz,Halimeh:2020ecg,VanDamme:2020rur,Nguyen:2021hyk,PhysRevA.104.062425,Halimeh:2021vzf,ARahman:2022tkr,Yeter-Aydeniz:2022vuy,Gustafson:2023swx} of QEM techniques has primarily focused on using zero noise extrapolation, random compilation, and readout error mitigation.
Because many approaches for digitization of LGTs (see Sec VI.b of~\cite{Bauer:2022hpo} for a review) have been proposed, it will be illuminating to compare their performance on currently accessible systems to understand which approaches should be pursued for large-scale quantum simulations.

Quantum electrodynamics (QED) in $1+1d$, also called the Schwinger model~\cite{Schwinger:1962tp}, demonstrates both chiral symmetry breaking and confinement. Therefore, the Schwinger model and its approximations provide relatively simple yet nontrivial tests cases for quantum simulation methods, and have been the focus of much activity. State-preparation techniques using adiabatic ~\cite{Kuhn:2014rha,Chakraborty:2020uhf} and  variational~\cite{Kokail:2018eiw,Xie:2022jgj,Yamamoto:2021vxp} methods as well as thermal-pure-quantum states~\cite{Davoudi:2022uzo} have been tested. Ground state properties have been calculated at finite temperature~\cite{Xie:2022jgj,deJong:2021wsd} and density~\cite{Yamamoto:2021vxp,Davoudi:2022uzo}, including topological terms~\cite{Magnifico:2019ulp,Chakraborty:2020uhf,Honda:2021aum}. Additional work has studied non-equilibrium dynamics~\cite{deJong:2021wsd,Kharzeev:2020kgc} and the role of dynamical quantum phase transitions~\cite{Mueller:2022xbg,Jensen:2022hyu}.
Error mitigation~\cite{Nguyen:2021hyk} and correction~\cite{Rajput:2021trn,Gustafson:2023swx} strategies have been developed for the Schwinger model, and practical resource estimates have been derived for some digitization schemes~\cite{Shaw:2020udc}.

In this work, we simulate $1+1d$ $\mathbb{Z}_2$ gauge theory with a single massive fermion, which is the simplest discrete subgroup approximation of the Schwinger model,~\cite{Magnifico:2019ulp,Bender:2018rdp,Mildenberger:2022jqr,Irmejs:2022gwv}\footnote{Other proposed schemes for rendering the photon-field Hilbert space finite dimensional include truncating the compact QED variables~\cite{Kuhn:2014rha}, quantum link models~\cite{Surace:2019dtp}, and quantum cellular automatons~\cite{Arrighi:2019edh}.} as well as a model of condensed matter Luttinger liquid systems with interesting phenomena that has attracted recent interest~\cite{Frank:2019jzv,Borla:2019chl,Borla:2020upy,Das:2021iwa,Irmejs:2022gwv,Samajdar:2022mtt, K_nig_2020,Yarloo_2019}. 
We compute real-time (Minkowski) two-point correlation functions and extract the energy of the lowest-lying spin-1 state from their time dependence.
In the infinite-volume, continuum limit this energy corresponds to the mass of the lightest spin-1 fermion-antifermion bound state in the theory.
In realistic four-dimensional LGT studies, a set of analogous quantities can be matched to experimentally measured particle masses in order to set the scale of the lattice spacing in physical units and fix the quark masses.  This scale setting is a necessary step before calculations of other spectral quantities can be used to predict other particle masses as well as some non-equilibrium observables.
Strategies for determining the lattice spacing in quantum simulations have been proposed in Refs.~\cite{Carena:2021ltu,Clemente:2022cka,Carena:2022hpz}.

The remainder of this paper is organized as follows.
We first discuss in Sec.~\ref{sec:theory} the Hamiltonian time evolution in $\mathbb{Z}_2$ gauge theory and its implementation via quantum circuits in Sec. \ref{sec:theory}. Sec.~\ref{sec:qem} discusses the study of QEM strategies for these circuits. We apply the QEM strategies in Sec.~\ref{sec:numresult} to compute the real-time evolution of fermion-antifermion bound states and determine the mass gap from these results. The performance and interplay of various QEM techniques is detailed for simulations using multiple IBM quantum computers. Finally, we conclude with a discussion of the results and future work in Sec.~\ref{sec:Conclusions}.

\section{Theory}
\label{sec:theory}

In this work we use the Kogut-Susskind lattice Hamiltonian for $1+1d$ $\mathbb{Z}_2$ gauge theory with two-component staggered fermionic matter and open boundary conditions (OBCs)~\cite{PhysRevD.11.395,PhysRevD.13.1043}.\footnote{Open boundary conditions minimize the complexity of our circuits on the available IBM hardware topology (see Sec.~\ref{sec:circuits}).}   
A convenient feature of this model is that both the fermionic and bosonic degrees of freedom have the same local Hilbert space dimension as qubits. This allows for a straightforward mapping to qubit-based quantum computers. Further simplifications are possible in $1+1d$ to reduce the quantum resources required, {\it e.g.,} integrating out the nondynamical gauge fields or using the block structure of the symmetry sectors. We do not employ such optimizations here, however, since they do not persist in higher dimensions~\cite{Zohar:2021nyc}.  

Our simulations were performed on the IBM quantum computers \manila, \nairobi, \jakarta, and \quito. Typical specifications for these machines during our running period are provided in the files accompanying this work, while their physical qubit layouts are shown in Fig.~\ref{fig:qpumapping}.

\subsection{Hamiltonian}\label{sec:hammy}

\begin{figure}
    \centering
    \includegraphics[width=\linewidth]{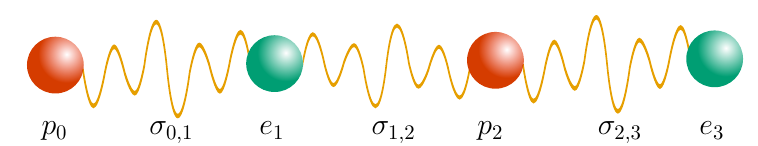}
    \caption{Pictorial representation of the $N_s=4$ one-dimensional lattice used in this work. The lattice is represented using seven qubits: two represent ``electron'' components of the staggered fermion $e_n$ (green), two represent the analogous ``positron'' components $p_n$ (red), and three represent $\mathbb{Z}_2$-valued ``photon'' fields for which operators are labeled by adjacent pairs of lattice sites, e.g. $\sigma_{n,n+1}$ (orange).}
    \label{fig:schlat}
\end{figure}

The Hamiltonian governing $\mathbb{Z}_2$ gauge theory on a lattice of length $N_s$ with open boundary conditions is
\begin{eqnarray}
    H &=& \sum_{n=1}^{N_s - 1}\Big[\frac{1}{2 a_s}\sigma^{x}_{n,n+1} + \frac{a_s \eta }{2}(\bar{\psi}_n\sigma_{n,n+1}^{z}\psi_{n+1} + h.c.)\Big]\nonumber\\
    && +\ a_s m_0 \sum_{n=1}^{N_s}  (-1) ^{n}\bar{\psi}_n\psi_n,
\end{eqnarray}
where $m_0$ and $\eta$ are the bare fermion mass and the gauge coupling respectively, $\psi_n$ is a fermion field on lattice site $n$, $\bar{\psi}_n$ is the corresponding antifermion field, and the Pauli matrices $\sigma_{n,n+1}^x$ and $\sigma_{n,n+1}^z$ act on the two-component state of the $\mathbb{Z}_2$ gauge-field, and $a_s$ is the spatial lattice spacing.   The three terms in $H$ represent the gauge kinetic term, the fermion hopping term, and the fermion mass term, respectively.  For the remainder of the paper, we will set $a_s = 1$ such that all dimensionful quantities are in lattice-spacing units.

In the Kogut-Susskind fermion formulation, a single 2-component spinor $\psi(x)$ is described by two 1-component staggered fields $\psi_n$ on neighboring sites. Here we denote the components on even sites by $p_n$ (for positron) and the components on odd sites by $e_n$ (for electron). The field content of an $N_s$-site lattice with OBCs is $N_s/2$ electron sites, $N_s/2$ positron sites, and $N_s-1$ gauge-field links where $N_s$ must be even. A depiction for the case of $N_s=4$ is provided in Fig.~\ref{fig:schlat}. 

After a Jordan-Wigner transformation~\cite{JordanPWignerE} to convert the fermionic degrees of freedom to bosonic ones,
$H$ becomes

\begin{equation}
    \begin{split}
\label{eq:z2ham}
H = & \frac{1}{2} \sum_{n=0}^{N_s - 1} \sigma^{x}_{n,n+1} - \frac{m_0}{2} \sum_{n=0}^{N_s - 1} (-1)^n Z_{n}\\
&  + \frac{\eta}{4} \sum_{n=0}^{N_s-2}(X_n X_{n + 1} + Y_{n} Y_{n + 1})\sigma^{z}_{n,n+1}.
\end{split}
\end{equation}
The operators $X_n$, $Y_n$, and $Z_n$ denote Pauli matrices acting on the electron and positron qubit states.
This form of the Hamiltonian is used in the quantum simulations below.

\subsection{Circuits}\label{sec:circuits}

\begin{figure}[tb!]
    \centering
    \includegraphics[width=0.7\linewidth]{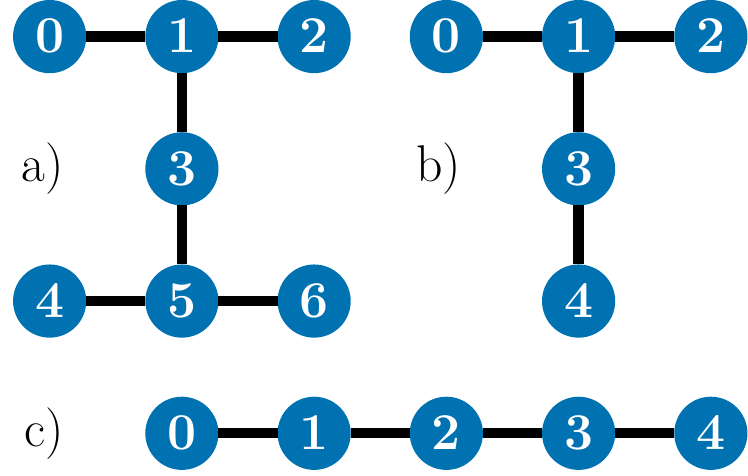}
    \caption{Qubit layout and connectivity for a) \nairobi\, and \jakarta, b) \manila, and c) \quito.}
    \label{fig:qpumapping}
\end{figure}

\begin{figure*}[tb!]
    \centering
\includegraphics[width=\linewidth]{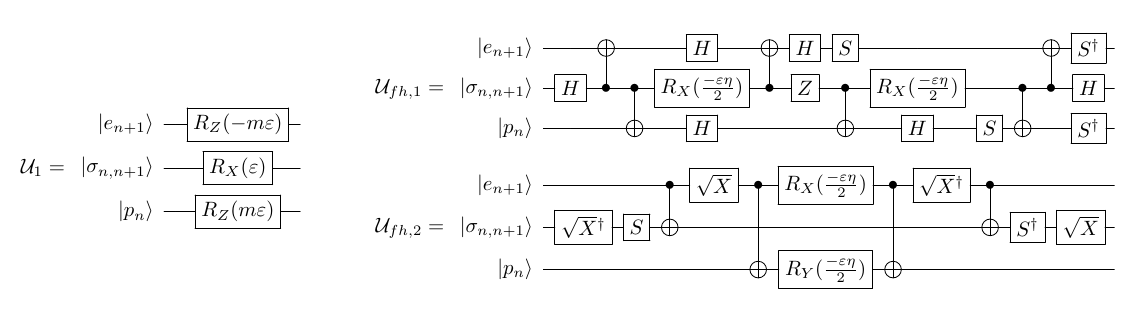}
\caption{The quantum circuit $\mathcal{U}_{1}$ implementing the gauge kinetic term and the fermion mass term is shown in the left panel. In the right panel, two equivalent quantum circuits implementing the fermion hopping term appearing in $\mathcal{U}_{2}$ and $\mathcal{U}_{3}$ are shown. These two circuits have different qubit connectivities and are used in conjunction to provide efficient mappings between logical and physical qubits as described in the main text.
\label{fig:circs}
}
\end{figure*}

For $\mathbb{Z}_2$ lattice gauge theory with Kogut-Susskind fermions, the state of each fermion component and gauge field can be encoded in a single qubit. Hence, a lattice with $N_s$ sites (or, equivalently, $N_x = N_s/2$ spatial points) can be encoded in $N_q = 2N_s - 1 = 4N_x - 1$ qubits.  Here we simulate 2-site and 4-site systems using 3 and 7 qubits, respectively.

We approximate the time evolution operator $U(t)=e^{-itH}$ via second order Trotterization~\cite{trotter1959product,Suzuki:1985,PhysRevX.11.011020} as,
\begin{equation}
    \label{eq:secondordertrotter}
    U(t) \approx \mathcal{U}(t/\varepsilon)^{N_t} ,
\end{equation}
where $N_t$ is an integer and $\epsilon \equiv t/N_t$.
Performing simulations of real-time evolution on a quantum computer requires encoding $\mathcal{U}(t/\varepsilon)$ as a quantum circuit.
To simplify the circuit encoding, we first express the Hamiltonian as a sum of three commuting operators
\begin{equation}\label{eq:H2andH3}
\begin{split}
H_1 &=  \frac{1}{2} \sum_{n=0}^{N_s - 1} \sigma^{x}_{n,n+1} - \frac{m_0}{2} \sum_{n=0}^{N_s - 1} (-1)^n Z_{n}, \\
H_2 &=  \frac{\eta}{2} \sum_{n=\text{even}}^{N_s-2}(X_n X_{n + 1} + Y_{n} Y_{n + 1})\sigma^{z}_{n,n+1},\\
H_3 &=  \frac{\eta}{2} \sum_{n=\text{odd}}^{N_s-2}(X_n X_{n + 1} + Y_{n} Y_{n + 1})\sigma^{z}_{n,n+1},
\end{split}
\end{equation}
each of which is easily diagonalizable. Operator $H_1$ comprises the gauge-kinetic and fermion-mass terms, while $H_2$ and $H_3$ are the even- and odd-site fermion-hopping terms, respectively. In terms of the $H_i$, the Hamiltonian in Eq.~(\ref{eq:z2ham}) is given by 
\begin{equation}
    H = \sum_{i=1}^3 H_i,
\end{equation}
and the Trotterized time-evolution operator in Eq.~(\ref{eq:secondordertrotter}) is 
\begin{equation}
    \mathcal{U}(t/\varepsilon) = \prod_{i=1}^3 \mathcal{U}_i(t/\varepsilon),
\end{equation}
where
\begin{equation}
\mathcal{U}_i(t) \equiv e^{-it H_i}.
\end{equation}

For the simulations described in this work, we time evolve the quantum system using the circuits shown in Fig.~\ref{fig:circs}. Operator $\mathcal{U}_1$ 
requires only single-qubit gates, and we employ the simple circuit implementation shown in the left panel. Operators $\mathcal{U}_{2}$ and $\mathcal{U}_{3}$ require entangling gates because the hopping term couples neighboring fermion and gauge degrees of freedom. They can be represented by the same quantum circuit; however, since $H_2$ and $H_3$ differ only by the lattice sites upon which they act. We employ the two equivalent circuit representations of the hopping-term operator in our simulations, $\mathcal{U}_{fh,1}$ and $\mathcal{U}_{fh,2}$.  These are shown in the upper- and lower-right panels, respectively. Because the locations of the entangling CNOT gates differ between $\mathcal{U}_{fh,1}$ and $\mathcal{U}_{fh,2}$, combining both circuit variations allows us to more efficiently use the available quantum hardware without the need for costly SWAP gates. 

For circuit $\mathcal{U}_{fh,1}$, in which the fermion and anti-fermion are coupled via a central gauge field, the physical qubit layout matches the 1$d$ lattice picture in Fig.~\ref{fig:schlat}. This straightforward implementation has been employed in recent quantum simulations of the lattice Schwinger model and its approximations e.g.~\cite{Chakraborty:2020uhf,Yamamoto:2021vxp,deJong:2021wsd}, and it is well-suited to quantum hardware with linear topology. Many NISQ-era computers, however, have nonlinear qubit layouts~\cite{ionq_2020,ibm_2023,rigetti_2023,google_2020}. On these devices, the number of lattice sites that can be simulated using the hopping-term circuit layout $\mathcal{U}_{fh,1}$ is limited by the longest linear graph, and the resulting efficiency loss can be substantial. For example, on the machines \jakarta\ and \nairobi\ (see Fig.~\ref{fig:qpumapping} a), with the standard qubit layout $\mathcal{U}_{fh,1}$ only 3 out of 7 qubits would usable, thereby wasting over 60\% of the available resources.

\begin{figure}[tb!]
    \centering
    \includegraphics[width=\linewidth]{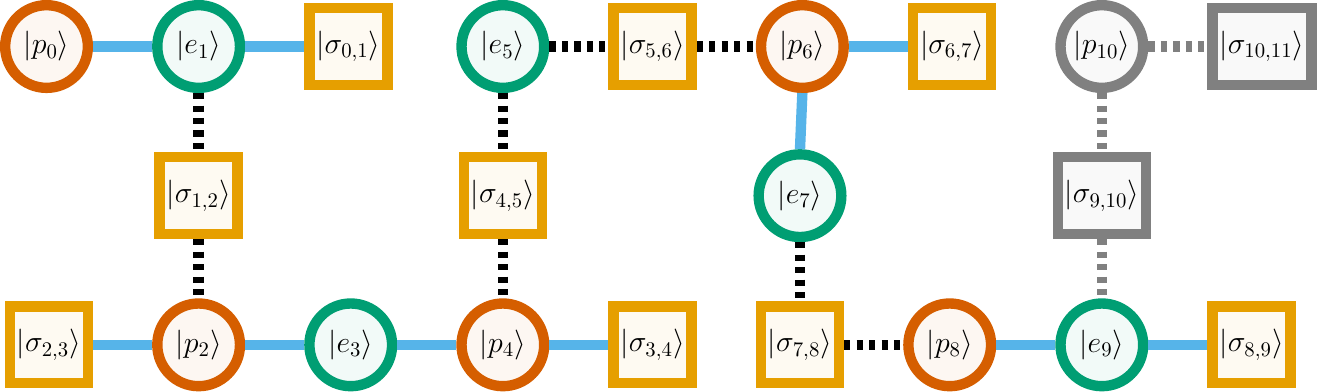}
    \caption{Example mapping of 1+1d $\mathbb{Z}_2$ which tessellates heavy-square qubit connectivity layouts relevant for \nairobi\ and \jakarta. Solid (dashed) lines indicate $U_{fh,1}$ ($U_{fh,2}$) gates are used to implement fermion hopping terms involving a given pair of lattice sites and grayed qubits denote ones unnecessary for even numbers of lattice sites. This is the fewest number of idle qubits possible for this graph.}
    \label{fig:perthmapping}
\end{figure}

To address this limitation, we employed circuit identities to derive from the standard circuit $\mathcal{U}_{fh,1}$ a new circuit $\mathcal{U}_{fh,2}$ in which the gauge and anti-fermion fields are coupled via a central fermion. Using both fermion hopping-term circuits together allows us to employ all 7 quibits on \jakarta\ and \nairobi\ in our quantum simulations. 
 Further, these two circuits form a basis for efficient 1+1d ${\mathbb Z}_2$ simulations using all qubits on any quantum device with heavy-polygon topology.\footnote{An $N$-sided heavy polygon has qubits on both the $N$ edges and the $N$ vertices.} To illustrate how $\mathcal{U}_{fh,1}$ and $\mathcal{U}_{fh,2}$ can be effectively combined, Fig.~\ref{fig:perthmapping} shows an example circuit mapping of 1+1d ${\mathbb Z}_2$ LGT with 5 spatial points onto a 22-qubit heavy-square device. This is the maximum possible efficiency since simulating this system with $N_x$ spatial points requires $4N_x - 1$ qubits as discussed above.

\subsection{Simulation Prescription}

Correlation functions describing the real-time evolution of a quantum state of interest $\ket{\phi} \equiv \phi^\dagger \ket{0}$ and subsequent interaction with a Hermitian\footnote{Computing the expectation value of non-Hermitian operators is possible but requires multiple circuits and ancilla qubits as in e.g. the Hadamard test (see Sec. 2.4.3 of Ref~\cite{quantumalgorithms}).} operator $O$ are generically of the form
\begin{equation}
  C(t) =  \langle\phi | U^\dagger(t) O U(t) | \phi \rangle=\langle 0|\phi \, U^\dagger(t) O U(t) \, \phi^\dagger | 0 \rangle. \label{eq:correlator}
\end{equation}
By inserting complete sets of energy eigenstates, it is straightforward to show that $C(t)$ has the spectral representation
\begin{equation}
  C(t) = \sum_{n,m} \braket{\phi}{E_m } \braket{ E_n }{ \phi } \mbraket{m}{O}{n} e^{-i(E_n - E_m)t}. \label{eq:spec}
\end{equation}
Measurements of $|C(t)|^2$ from quantum simulations can be fit to this oscillatory form in order to extract energy differences $E_n - E_m$.

\begin{figure}[!t]
\includegraphics[width=0.6\linewidth]{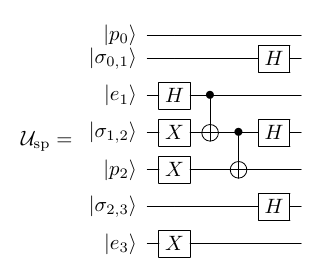}
\caption{Quantum circuit implementing initial-state preparation as $|\psi(4)\rangle = \mathcal{U}_{\text{sp}}|0....0\rangle$.}
\label{fig:stateprep}
\end{figure}

Because we do not explicitly project to the gauge-invariant sector of Hilbert space during time evolution, our simulations must start with gauge-invariant initial states in order to construct gauge-invariant correlation functions.
Here we choose our initial state to be a linear superposition of two gauge-invariant states: the non-interacting vacuum state $\ket{\Omega}$  and a state $\ket{P}$ that is expected to have significant overlap with excited states such as electron-positron bound states.
Explicitly, the states $|\Omega(N_s)\rangle$ and $|P(N_s)\rangle$ are defined for lattices of size $N_s$ as
\begin{equation}
  \begin{split}
    \label{eq:compstates}
    |\Omega(N_s)\rangle = &\left(\prod_{n=\text{even}}^{N_2-2}H_{n,n+1}X_{n+1}\right)\ket{0}^{\otimes 2N_s-1}, \\
    |P(N_s)\rangle = &X_m\sigma_{m,m+1}X_{n+1}\ket{\Omega(N_s)}.
  \end{split}
\end{equation}
where $H$ is the Hadamard gate and $m=N_s/2-1$ is the center lattice site.
The superposition of these two states
\begin{equation}
    \label{eq:initstate}
    \ket{\phi(N_s)} = \frac{1}{\sqrt{2}}\Big(|\Omega(N_s)\rangle + |P(N_s)\rangle\Big),
\end{equation}
is used as the initial state in our simulations.
The circuit to build this state from the computational $|0....0\rangle$ state for $N_s=4$ is given in Fig. \ref{fig:stateprep}.

In this work, we compute correlation functions involving this state and the operator
\begin{equation}
\label{eq:exciteoperator}
    O = \psi^{\dagger}_n\sigma^{z}_{n,n+1}\psi_{n+1} + \psi^{\dagger}_n \sigma^{z}_{n, n+1}\psi^\dagger_{n+1} + h.c.\text{ .}
\end{equation}
This operator takes a simple form in the qubit spin basis:
\begin{equation}
  O = X_n \sigma^z_{n,n+1} X_{n+1},
\end{equation}
which allows it to be included efficiently in quantum simulations.
Further, using Eq.~\eqref{eq:compstates} shows that diagonal matrix elements of $O$ vanish for these states,
\begin{equation}
  \mbraket{ \Omega(N_s) }{ O }{ \Omega(N_s) } = 0 = \mbraket{ P(N_s) }{ O }{ P(N_s) },
\end{equation}
while off-diagonal matrix elements are unity,
\begin{equation}
  \mbraket{ \Omega(N_s) }{ O }{ P(N_s) } = 1 = \mbraket{ \Omega(N_s) }{ O }{ P(N_s) }.
\end{equation}
Thus, the value of $C(t=0)$ can be computed exactly:
\begin{equation}
  \begin{split}
    C(0) &= \mbraket{ \phi(N_s) }{ O }{ \phi(N_s) } = 1.
  \end{split}
\end{equation}

Correlation functions $C(t)$ defined using this state as in Eq.~\eqref{eq:correlator} have time-dependence generically given by Eq.~\eqref{eq:spec}, and in the limit of large $t$ their Fourier transformations should be sharply peaked about values corresponding to some energy differences.
If $\ket{\Omega}$ and $\ket{P}$ each overlap predominantly with a single energy eigenstate with energy $E_{\Omega}$ and $E_P$, respectively, then the time-dependence of $C(t)$ reduces to the simple form  
\begin{equation}
\begin{split}
  \mbraket{ \phi }{ U^\dagger(t) O U(t) }{ \phi } &\approx \cos[(E_P-E_{\Omega})t] \mbraket{\Omega}{O}{P} \\
  &\hspace{10pt} + \frac{1}{2} \mbraket{\Omega}{O}{\Omega} + \frac{1}{2} \mbraket{P}{O}{P} \\
  &= \cos[(E_P-E_{\Omega})t]. \label{eq:spec01}
  \end{split}
\end{equation}
This means that $C(t)$ can be expressed as
\begin{equation}
  \begin{split}
    C(t) &= \mbraket{ \phi(N_s) }{ U^{\dagger}(t) \ O \  U(t) }{ \phi(N_s) } \\
    &=\cos(M t) + \ldots, \label{eq:cos}
  \end{split}
\end{equation}
where $M$ is the energy difference between the eigenstates dominantly overlapping with $\ket{P(N_s)}$ and $\ket{\Omega(N_s)}$ and the $\ldots$ denotes contributions from other states that can lead to different $t$-dependence than a single-cosine form. 
Fits of quantum simulation results to Eq.~\eqref{eq:cos} can be used to study the validity of this approximation.
If the single-state contribution $\cos(Mt)$ provides a good fit to correlation function results, then the fit parameter $M$ can be expected to describe the energy gap between an electron-positron bound state and the vacuum, which corresponds to the mass of the particle associated with this state in the continuum and infinite-volume limits.
This accuracy of this method can be directly tested by comparing fitted results for $M$ with exact spectral results, which can be numerically computed on classical computers for small $N_s$.

Finally, we employ the Trotterized approximation to $C(t)$ in our simulations, which is defined as
\begin{equation}
    \label{eq:simobservable}
    \mathfrak{C}(t/\varepsilon) = \mbraket{ \phi(N_s) }{ \mathcal{U}^{\dagger}(t/\varepsilon)^{N_t} \ O \  \mathcal{U}(t/\varepsilon)^{N_t} }{ \phi(N_s) },
\end{equation}
where $\mathcal{U}(t/\varepsilon)$ is defined in Eq.~\eqref{eq:secondordertrotter}.

\begin{table}
\caption{Details of the simulations performed in this work. The different dynamic decoupling (DD) schemes are described below in Sec.~\ref{subsec:DD}.}
\label{tab:datainfo}
\begin{tabular}{cccclcc}
\hline\hline
     $m_0$ & $N_s$ & Machine & Date & Time\footnote{Central Daylight Time} & DD & Qubits\footnote{Layouts found in Fig.~\ref{fig:qpumapping}}\\
     \hline
     1 & 2 & \jakarta & 1/3/23 & 20:59\footnote{Random compilation not used} &  None & (5,6,3)\\
     1 & 2 & \jakarta & 1/4/23 & 18:41\footnotemark[3] &  XY4 & (5,6,3)\\
     1 & 2 & \nairobi & 9/10/22 & 13:28 &  None & (5,4,6)\\
     1 & 2 & \nairobi & 9/10/22 & 11:53 &  XY4 &  (5,4,6)\\
     1 & 4 & \jakarta & 9/21/22& 15:23 & None &  (0,2,1,3,5,4,6)\\
     1 & 4 & \jakarta & 9/21/22 & 19:39 & XY4 &  (0,2,1,3,5,4,6)\\
     1 & 4 & \nairobi & 1/5/23& 08:57\footnotemark[3] & None& (0,2,1,3,5,4,6)\\
     1 & 4 & \nairobi & 1/5/23& 09:30\footnotemark[3] & XY4 &  (0,2,1,3,5,4,6)\\
     2 & 2 & \jakarta & 8/27/22 & 22:04 & None  & (5,6,3)\\
     2 & 2 & \jakarta & 9/22/22 & 21:44 & XY4  & (5,6,3)\\
     2 & 2 & \nairobi & 1/4/23 & 21:41\footnotemark[3]  & None & (5,4,6)\\
     2 & 2 & \nairobi & 1/5/23 & 22:51\footnotemark[3]  & XY4 & (5,4,6)\\
     2 & 4 & \jakarta & 1/4/23 & 20:36\footnotemark[3] & None & (0,2,1,3,5,4,6)\\
     2 & 4 & \jakarta & 1/4/23 & 21:01\footnotemark[3] & XY4 & (0,2,1,3,5,4,6)\\
     2 & 4 & \nairobi & 9/12/22 & 11:26 & None & (0,2,1,3,5,4,6)\\
     2 & 4 & \nairobi & 9/14/22 & 11:22 & XY4 & (0,2,1,3,5,4,6)\\
     \hline
     1 & 2 & \manila & 8/4/22& 10:25 &XY4 & (3,1,4)\\
     1 & 2 & \manila & 8/4/22& 09:06 & CMPG  & (3,1,4)\\
     1 & 2 & \manila & 8/4/22& 10:26 & EDD & (3,1,4)\\
     1 & 2 & \manila & 8/4/22& 16:57 &XY4& (1,0,2)\\
     1 & 2 & \manila & 8/4/22& 14:58 & CMPG & (1,0,2)\\
     1 & 2 & \manila & 8/4/22& 16:56 & EDD&  (1,0,2)\\
     1 & 2 & \quito & 7/27/22 & 13:38 & XY4& (3,2,4)\\
     1 & 2 & \quito & 7/27/22 & 13:26 & CMPG & (3,2,4)\\
     1 & 2 & \quito & 7/27/22 & 13:39 & EDD& (3,2,4)\\
     1 & 2 & \quito & 7/27/22 & 12:05 & XY4& (1,0,2)\\
     1 & 2 & \quito & 7/27/22 & 11:46 & CMPG & (1,0,2)\\
     1 & 2 & \quito & 7/27/22 & 12:08 & EDD& (1,0,2)\\
     \hline\hline
\end{tabular}
\end{table}

\label{sec:spec}

\section{Error Mitigation Of a Quantum Simulation}
\label{sec:qem}
Many quantum error mitigation strategies have been studied for reducing the systematic uncertainties associated with errors in NISQ era quantum simulations. 
A primary goal of this work is to study the interplay between different QEM methods and the reliability of quantum simulation results using combinations of state-of-the-art methods.
This section briefly introduces the QEM methods that we found to provide significant improvements in these calculations: randomized compiling, readout error mitigation, rescaling, and dynamic decoupling.

These QEM strategies introduce additional correlations between quantum simulation results, and it is important to accurately determine and include these correlations in analyses of quantum simulation results.
Throughout this work we use bootstrap resampling~\cite{10.1214/aos/1176344552,davison_hinkley_1997,young_stats} to determine all statistical uncertainties and correlations between observables, in particular using correlated resampling of quantities arising through QEM that appear in multiple observables computed using the same quantum computer.

We performed quantum simulations of $\mathfrak{C}(t)$ using the parameter choices $m_0 \in \{1,2\}$ and $N_s \in \{2,4\}$ in the Hamiltonian in Eq. (\ref{eq:z2ham}) with $\eta=1$ in all cases.
Each simulation was performed for $N_t=20$ Trotter steps with $\varepsilon=0.3$.  For each $N_t$, $N_{\text{rc}}=30$ randomly compiled circuits were run with $N_{\text{meas}}=2,000$ measurements collected for each.
These production simulations were carried out on \nairobi~and \jakarta~while additional testing simulations were also investigated on \manila~and \quito.
The full details of the simulations are listed in Table \ref{tab:datainfo}.
For each simulation, we performed $N_{\text{meas}} = N_{\text{shots}} N_{\text{rc}} = 6 \times 10^4$ measurements of each quantum circuit and used $N_{\text{boot}} = 10^4$ random bootstrap samples of this ensemble of measurements. Bootstrap covariance matrices were determined for all correlated observables and the associated uncertainties were propagated to fitted quantities in a correlated way using the \texttt{gvar} and \texttt{lsqfit} packages \cite{Lepage:2001ym,Bouchard:2014ypa,McNeile:2010ji,Hornbostel:2011hu,Dowdall:2019bea}.

\subsection{Randomized Compiling}

\begin{figure*}[!t]
\includegraphics[width=\linewidth]{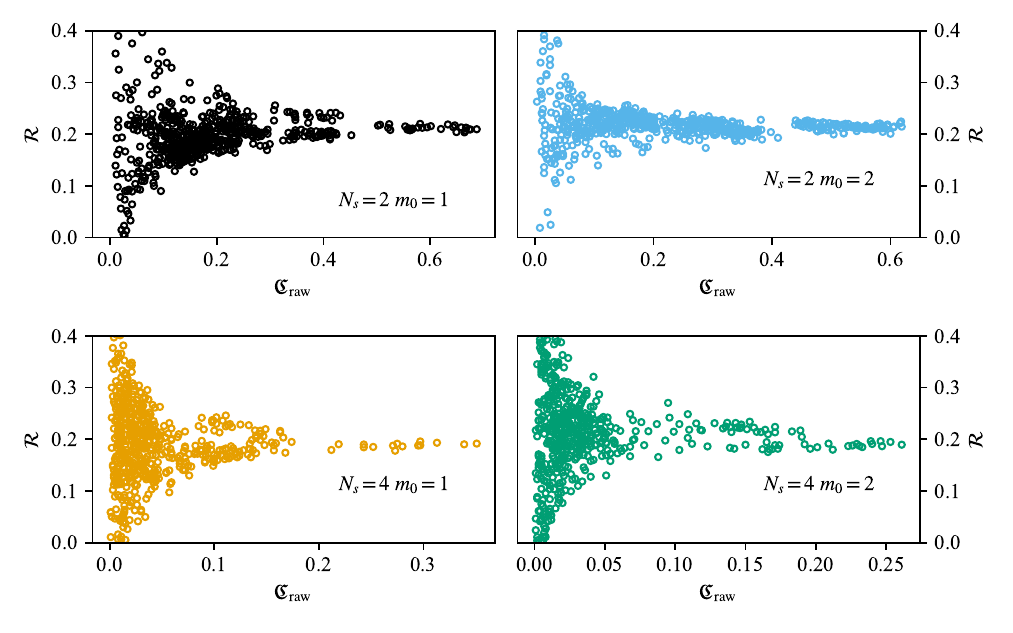}
\caption{ Relative  shifts $\mathcal{R}$ in the Trotterized correlation function $\mathfrak{C}_{\text{raw}}$ (aggregated across $t/\varepsilon$) defined in Eq.~(\ref{eq:relativeshift}) due to readout mitigation for \nairobi\, and \jakarta. The data points are aggregated across all possible randomized compiling circuits that did not use dynamic decoupling. Each data point was measured with $N_{\text{meas}} = 2,000$.}
\label{fig:readoutversusmachine}
\end{figure*}

Randomized compiling (RC) transforms coherent systematic uncertainties associated with the imperfect fidelity of quantum gates into stochastic systematic uncertainties that can be quantified with a Markovian noise model
~\cite{PhysRevA.94.052325,Erhard_2019,li2017efficient, 2018efficienttwirling, 2013PhRvA..88a2314G, 2016efficienttwirling,Silva-PT2008,Winick:2022scr}. It has seen great success in other $(1+1)d$ LGT applications~\cite{Yeter-Aydeniz:2022vuy,Carena:2022kpg,Farrell:2022wyt,Gustafson:2022xdt,Atas:2022dqm}. At present it must be implemented by hand, but it is expected to become a standard part of transpilation through parametric compilation~\cite{Abrams2020,PhysRevA.97.022330,PhysRevApplied.10.034050,Benedetti_2019,PhysRevResearch.2.033125,Katabarwa:2021vdk,PhysRevResearch.2.033447}.
RC utilizes multiple gates that are equivalent in the absence of noise but that differ in the presence of gate errors.
Therefore, the averages of the results often have smaller errors than any individual circuit and the variance of the results provides a partial measure of the size of systematic uncertainties arising from gate errors.

A strategy for implementing randomized compiling is ``Pauli twirling'', in which a gate $\Lambda$ is replaced by a gate including additional sets of Pauli gates $\lbrace \sigma_i\rbrace$  and $\lbrace \sigma'_i\rbrace$, where $i$ indexes the qubits acted on by $\Lambda$, that are chosen to satisfy~\cite{Silva-PT2008,2013PhRvA..88a2314G}
\begin{equation}
    \label{eq:paulicomp}
    \left[ \bigotimes_{i} \sigma_i  \right] \Lambda  \left[ \bigotimes_{j} \sigma_j'   \right] = \Lambda.
\end{equation}
Any solution to Eq.~\eqref{eq:paulicomp} provides a valid gate that is equivalent to $\Lambda$ on an ideal quantum computer and can be used for randomized compiling.
For any choice of $\lbrace \sigma_i\rbrace$, the $\lbrace \sigma'_i\rbrace$ required to produce such a solution is simply obtained by multiplying both sides Eq.~\eqref{eq:paulicomp} by the inverses of the gates appearing and is given by
\begin{equation}
    \label{eq:paulicomp2}
      \left[ \bigotimes_{j} \sigma_j'   \right] = \Lambda^\dagger \left[ \bigotimes_{i} \sigma_i  \right] \Lambda.
\end{equation}
For complicated gates such as the fermion hopping term above, 64 solutions to Eq.~\eqref{eq:paulicomp2} can be produced in this manner, while for the CNOT there are 16.
In order to better handle circuit scheduling constraints on cloud computing platforms, a random set of $N_{\rm RC}$ solutions can be chosen at compile time, with solutions chosen independently for each instance of $\Lambda$ appearing in a quantum circuit.
The optimal value of $N_{\rm twirl}$ depends on both $\Lambda$ and the hardware is run and can be determined empirically by increasing $N_{\rm twirl}$ until the effects of randomized compiling saturate or are offset by a prohibitively larger number of simulations.

Pauli twirling removes correlations between repeated $\Lambda$, but any internal correlations persist.  Thus as the correlation between native gates decreases with hardware and implementation improvements, resources devoted to Pauli twirling can be reduced by only implementing them for larger gates like $U_{fh}$.  On the present systems, we investigated Pauli twirling at the level of $U_{fh}$ and at the level of the CNOTs within it. A mild but statistically significant preference to twirling the CNOTs was observed which will be used for the remainder of this work.
\subsection{Readout Error Mitigation}

The measurement operation on quantum computers is quite noisy. There are many causes for these errors such as classical bit-flips, amplitude dampening, and cross talk~ \cite{PhysRevApplied.14.054059,PhysRevApplied.12.054023,PhysRevApplied.10.034040,Sarovar2020detectingcrosstalk,Berg:2020ibi,Smith:2021iwt,Rudinger:2021nhd}. It is important to mitigate these errors as they will bias the observed value of an operator measured on a quantum computer. 

\begin{figure}[t!]
\centering
\subfigure[$N_s=2$]{\includegraphics[width=\linewidth]{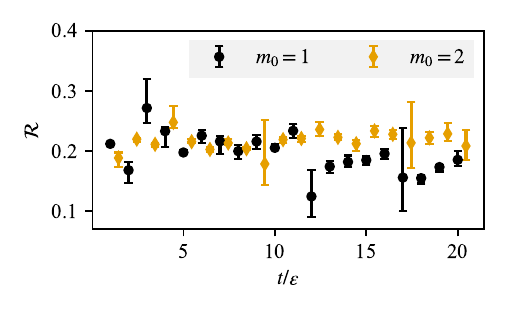}}
\subfigure[$N_s=4$]{\includegraphics[width=\linewidth]{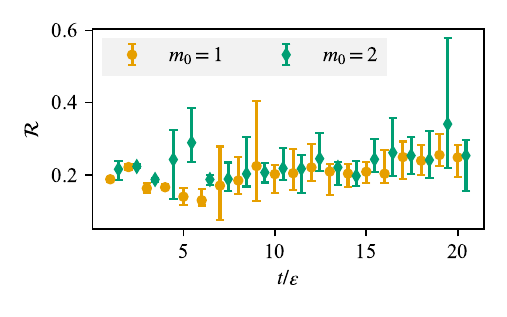}}
\caption{Effects of readout correction for various depth circuits corresponding to $t/\varepsilon$ Trotter steps. The top (bottom) figure corresponds to the $N_s = 2$ ($N_s = 4$) lattice volume, and the different colored points in each plot correspond to the different bare fermion masses indicated. Note that simulations for different masses were performed using different machines in both cases (\nairobi~ for parameters $\{N_s=2,m_0=1\}$ and $\{N_s=4,m_0=2\}$ and \jakarta~ for the other combinations), and the central values and uncertainties of the results will therefore differ. }. 
\label{fig:readoutwithdepth}
\end{figure}

While many methods to correct these errors exist~\cite{10.1145/3352460.3358265,Harrigan2021,PhysRevA.100.052315,Maciejewski2020mitigationofreadout,Nachman2020,Hicks:2021uvm}, we use  regularized response matrix inversion~\cite{PhysRevLett.122.110501,PhysRevA.101.032343,Hamilton:2020xpx,Geller_2021,PhysRevLett.119.180511}. For a single qubit if we prepare the system in the state $\ket{0}$ there is a probability $p_0$ that we measure the qubit in the state $\ket{0}$ and  a probability $1-p_0$ that we measure it in the $\ket{1}$ state. 
We can then use these and the analogous probabilities for an initial $\ket{1}$ state to construct a calibration matrix,
\begin{equation}
M = 
\begin{pmatrix}
p_0 & 1 - p_0\\
1 - p_1 & p_1\\
\end{pmatrix}.
\end{equation}
By acting on the vector of measured qubit state results with $M^{-1}$, one can mitigate readout error and return a ``corrected" output closer to the underlying distribution. Assuming that readout errors are uncorrelated, we can construct a tensor product $M^{\otimes N}$ and correct the readouts individually rather than the exponentially time consuming task of measuring all possible elements of the full readout correction matrix. \textsc{Qiskit RunTime} has readout mitigation built in~\cite{runtime}.

\begin{figure}[t!]
\includegraphics[width=\linewidth]{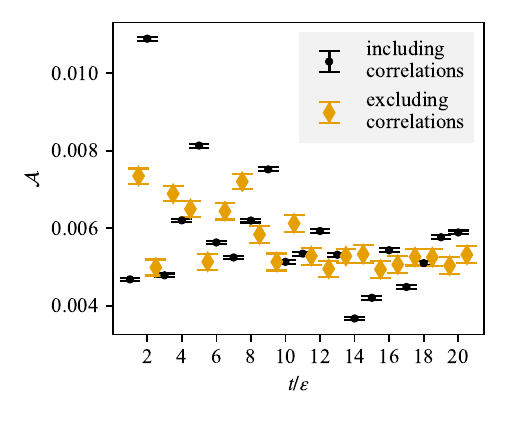}
\caption{Comparison of the absolute shifts from readout mitigation using calibration matrices computed with and without taking into account correlations between different circuit measurements.} 
\label{fig:correlatedaveragecomparison}
\includegraphics[width=\linewidth]{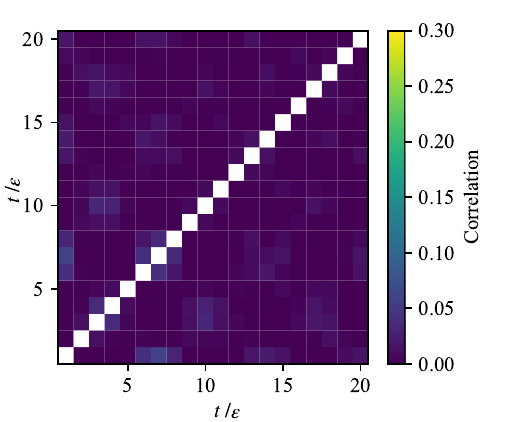}
\caption{Correlation matrices (that is, normalized covariance matrices) for circuits with different numbers of Trotter steps for the representative case of simulations with $m_0=1$ and $N_s=2$.}
\label{fig:correlationmatrixdiftime}
\end{figure}

The size of readout mitigation effects will depend on the observable under study, which is taken to be $\mathfrak{C}(t)$ below. 
The calibration matrix may introduce correlations between any simulations that are transformed using the same estimated process.
The absolute shift in the correlation function due to readout mitigation is defined as
\begin{equation}
    \label{eq:absoluteshift}
    \mathcal{A}(t) = |\mathfrak{C}_{\text{RO}}(t) - \mathfrak{C}_{\text{raw}}(t)|,
\end{equation}
where $\mathfrak{C}_{\text{RO}}(t) $ is the readout mitigated observable and $\mathfrak{C}_{\text{raw}}(t)$ is the observable calculated using the unmitigated data.
The relative shift is defined as 
\begin{equation}
\label{eq:relativeshift}
\mathcal{R}(t) = \frac{\mathcal{A}(t)}{| \mathfrak{C}_{\text{raw}}(t)|}.
\end{equation}
Figure \ref{fig:readoutversusmachine} shows the relative shift of the observable as a function of the observable magnitude for the simulations performed on \nairobi. 
The relative independence of $\mathcal{R}$ on $\mathfrak{C}_{\text{raw}}$ indicates that the absolute size $\mathcal{A}$ of readout mitigation effects is correlated with, and in particular approximately proportional to, $\mathfrak{C}_{\text{raw}}$.
This is not unexpected as errors on physical hardware are commonly asymmetric~\cite{PhysRevA.75.032345}.
Similar patterns were observed on \jakarta~ and together we can conclude the relative shift is approximately constant with circuit depth  as seen in Fig. \ref{fig:readoutwithdepth}.

Separate simulations with the same or different numbers of Trotter steps may become correlated because of the calibration matrix. 
The correlations between randomly compiled circuits were observed to be $\lesssim5\%$. Including these correlations has a noticeable effect when averaging these circuits, as seen in Fig. \ref{fig:correlatedaveragecomparison}.

Figure \ref{fig:correlationmatrixdiftime} shows the correlations introduced by readout mitigation between different time steps for the same parameters as Fig. \ref{fig:correlatedaveragecomparison}. It is unsurprising that observables involving different numbers of Trotter steps are less significantly correlated than those with the same number because the wave function is not as similar. 

\subsection{Rescaling}\label{sec:rescaling}
Measured observables will include an exponential decay with respect to the circuit depth~\cite{nielsen_chuang_2010}. It is possible to counteract this signal damping using rescaling~\cite{Urbanek2021,Vovrosh:2021ocf,ARahman:2022tkr}.
This method rescales the measurements of one set of circuits using information from a related set of circuits. The first set is the randomly compiled quantum simulation circuits $ \mathfrak{C}(t)$ of interest. The second set uses the same circuits except all non-Clifford gates are removed, which is denoted $r(t)$. The first set of circuits has an unknown output that the quantum simulation is designed to determine. On the other hand, the second set containing only Clifford gates can be efficiently simulated classically~\cite{Gottesman:1998hu}. Such classical algorithms can be extended further to some cases where some non-Clifford gates are allowed~\cite{PhysRevLett.116.250501,Bravyi2019simulationofquantum}.  This allows for comparison between the exact answer and the noisy result which can be used to mitigate some errors.

\begin{figure}
    \centering
    \includegraphics[width=\linewidth]{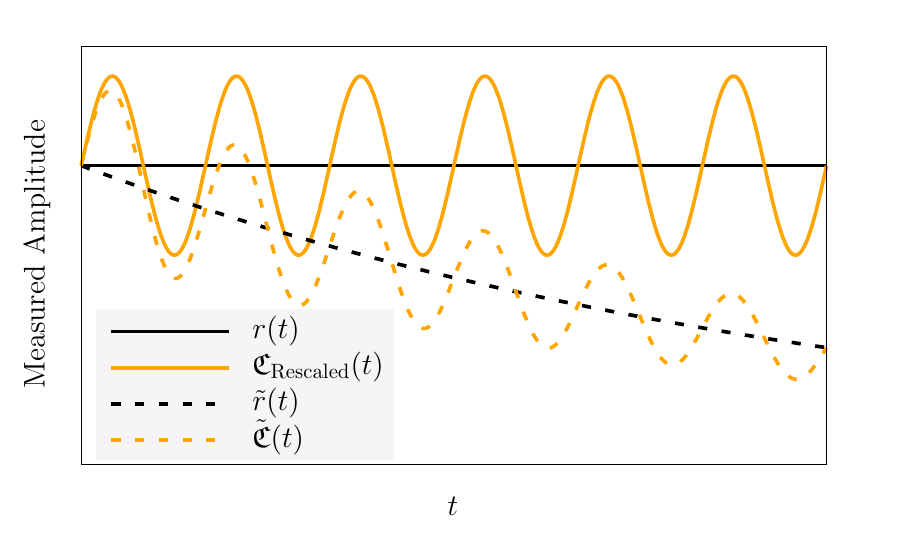}
    \caption{Pictorial representation of how rescaling circuits restore the signal. A noisy measurement (black dashed line) of an observable with a known $t$-independent expectation value (black solid line) is used to rescale a noisy measurement of an observable (yellow dashed line), resulting in the rescaled observable (yellow solid line).}
    \label{fig:rescale_cart}
\end{figure}

On a quantum computer, assuming only a depolarizing noise channel, the noisy estimate $\widetilde{r}(t)$ of the classically computable observable $r(t)$ is,
\begin{equation}
\label{eq:depolarizingrescale}
    \widetilde{r}(t) = (1 - \varepsilon)r(t) + \frac{\varepsilon}{2^n}\, \langle\text{Tr}[U_r(t)]\rangle,
\end{equation}
where $U_r(t)$ is an operator satisfying $r(t) = \langle U_r(t) \rangle$ for expectation values taken in the state $|\phi(N_s)\rangle$ and $\varepsilon$ is the strength of the depolarizing noise. 
Since $U_r(t)$ in our case is traceless and Hermitian because it is a tensor product of Pauli matrices, the result simplifies to $\widetilde{r}(t) = (1-\varepsilon)r(t)$.
Since $r(t)$ is easily computable, we can determine $(1-\varepsilon)$ from a measurement of $\widetilde{r}(t)$. Then we can correct for the same depolarizing noise in our correlation function of interest by rescaling the analogous noisy estimator $\widetilde{\mathfrak{C}}(t)$ by $(1-\varepsilon)^{-1}$~\cite{Urbanek2021,Vovrosh:2021ocf,ARahman:2022tkr}.
The resulting rescaled correlation function is given by
\begin{equation}
    \mathfrak{C}_{\text{Rescaled}}(t) = \frac{\widetilde{\mathfrak{C}}(t)}{1-\varepsilon} = \frac{\widetilde{\mathfrak{C}}(t) r(t)}{\widetilde{r}(t)}.
\end{equation}
This method is less feasible for long depth circuits because $\widetilde{r}(t)$ can be vanishingly small.  A pictorial representation is shown in Fig. \ref{fig:rescale_cart}.
The efficacy of rescaling is found to depend significantly on whether or not dynamical decoupling is included and is discussed further in the next section.

\subsection{Dynamical Decoupling}\label{subsec:DD} 

Dynamical decoupling (DD) is a method to reduce errors arising from spectator qubits that are acted on trivially by a given gate. When a qubit is idling, a set of single qubit operations are interleaved using basis transformations so that environmental contamination or spurious signals from other qubits become decoupled. As a result the coherence time of the quantum circuit becomes extended. 
For a review of the method see Ref. \cite{PhysRevLett.82.2417}.
There exist many methods for DD \cite{PhysRevLett.82.2417, Ezzell2022,Carr1954,Meiboom1958,Maudsley1986,Viola2003} and extensive studies on different DD sequences have been done \cite{Wrikat2019,Ezzell2022,Qi:2022gdn,https://doi.org/10.48550/arxiv.2209.05509}. It is well known the effectiveness of a given DD sequence is problem- \cite{Jurcevic_2021,Niu:2022wpa,Niu:2022jnx,Mundada:2022roq} and hardware-dependent \cite{Wrikat2019,Ezzell2022}.

\begin{figure*}
\includegraphics[width=\textwidth]{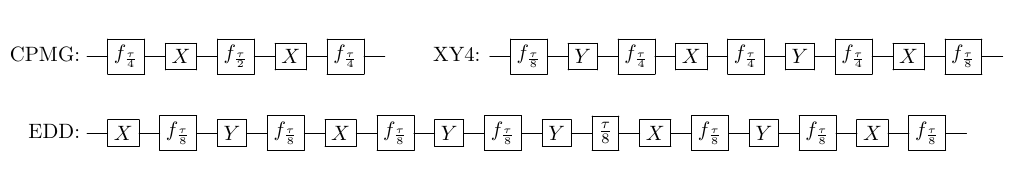}
\caption{The dynamical decoupling sequences studied in this work. The notation $\frac{\tau}{N}$ indicates a delay on the qubit equal to $\frac{\tau}{N}$. }
\label{fig:dynamicdecouplingsequences}
\end{figure*}

The time-dependent evolution can be described using a total Hamiltonian $H$ that depends on the Hamiltonian of the ideal system $H_{S}$, the Hamiltonian of the environment $H_{E}$, and the interaction between the system and the environment  $H_{SE}$ as
\begin{equation} \label{sys-env}
    H = H_{S}+H_{E}+H_{SE}.
\end{equation}
We can view $H_{SE}$ as an error term in the desired error-mitigated Hamiltonian and cancel them out to some degree using time-dependent inversion pulses in long periods of system-environment interaction.  These pulses can be incorporated into a circuit via a transpiler pass, as described below, and are available within \textsc{Qiskit}.

When a circuit is prepared to be run on a quantum computer, it is first transformed into a logically equivalent circuit in terms of the basis gates supported by the quantum computer through a process called transpiling. The DD transpiler pass \cite{DDqiskitdoc} analyzes a transpiled circuit for idle periods and inserts delay instructions. Although this will be effective in keeping a system in phase during single-qubit gates, CNOTs have longer gate times (10 or more times that of a single-qubit gate) and require a DD pulse sequence to decouple the idle qubits. Research on CNOT-induced idle periods and the best strategies for DD implementation is detailed in Ref.~\cite{Niu:2022IBM}.

In this work, we studied three different dynamical decoupling sequences: Carr-Purcell-Meiboom-Gill (CPMG)~\cite{Carr1954,Meiboom1958}, XY4~\cite{Maudsley1986}, and Eulerian dynamical decoupling (EDD)~\cite{Viola2003}, which are described below. These pulse sequences are shown in Fig. \ref{fig:dynamicdecouplingsequences} where, $\tau$ is the length of idle time on the qubit minus the single qubit gate operation times.

One of the earliest described decoupling sequences was proposed by Carr \& Purcell in 1954 \cite{Carr1954} and elaborated on by Meiboom \& Gill four years later \cite{Meiboom1958}. Called CPMG, this sequence gives first-order protection to environmental coupling. It involves two $X$ pulses, symmetrically placed on a spectator like in Fig.~\ref{fig:dynamicdecouplingsequences}.
While CPMG has been shown to perform better than a system without DD \cite{Ezzell2022}, it makes assumptions about the pulses being ideal. In addition, CPMG can only decouple states close to the equator of the Bloch sphere (such as the $|+\rangle$ and $|-\rangle$ states). To protect all states universally, more than just $X$ pulses are needed.

Following previous work on DD sequences, Maudsley brought forward a sequence with universal first-order protection of states in the ideal pulse limit \cite{Maudsley1986} by introducing a Y-rotation to the CPMG sequence as seen in Fig.~\ref{fig:dynamicdecouplingsequences}. 
This additional direction of rotation cancels out the final undesirable $H_{SE}$ terms.
It is important to note that to properly implement XY4, the total delay time $t$ must be bounded by the number of single qubit gates.
If the total CNOT time is less than the time it takes to implement 4 single qubit gates, then XY4 cannot be implemented. For IBM devices, this issue does not arise because CNOT times are 10 or more times that of single qubit gates. However, other factors such as pulse alignment can place restrictions on the allowed $f_{t}$. Like CPMG, XY4's efficacy is based on an ideal pulse model and may not accurately describe some noise sources arising on real quantum computers.

Eulerian Dynamical Decoupling (EDD) is a class of sequences proposed by Viola and Knill \cite{Viola2003} that provides universal first-order protection and takes into account imperfect pulses as well. The name of this procedure is derived from the Eulerian cycles on a Cayley graph of the discrete group (in our case, rotational gates from which a sequence is constructed), a formalism which is elaborated on in Ref.~\cite{Wrikat2019}. The explicit sequence of gates we used is described in Ref.~\cite{Ezzell2022} and shown in Fig.~\ref{fig:dynamicdecouplingsequences}.

\begin{figure}
\subfigure[Rescaling circuit observables]{
\includegraphics[width=\linewidth]{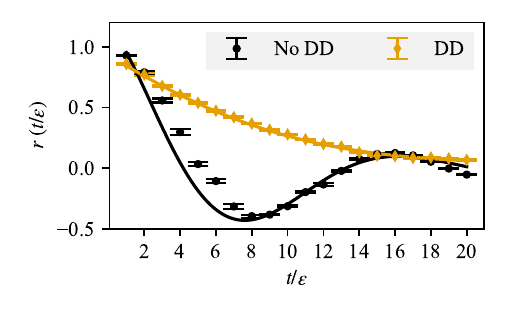}}
\subfigure[Evolution circuit observables]{
\includegraphics[width=\linewidth]{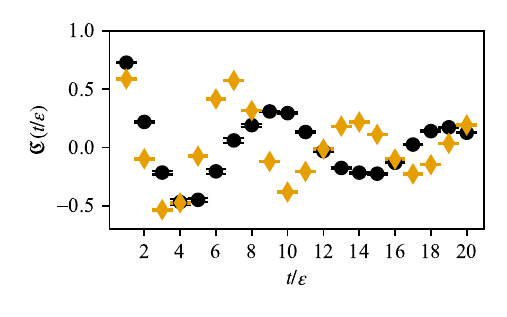}}
\caption{Effects on the rescaling circuit for $m_0=1$ and 3 qubits with and without using dynamical decoupling. The rescaling circuits $r(t)$ (or in the notation of Sec.~\ref{sec:rescaling} our noisy estimators $\widetilde{r}(t)$) for both cases are shown as a function of number of Trotter steps (top) along with the resulting (noisy estimator $\widetilde{\mathfrak{C}}(t)$ of the) unrescaled correlation function $\mathfrak{C}(t)$ (bottom) including readout mitigation but not rescaling.}
\label{fig:dynamicdecouplingeffectsrescale}
\end{figure}

Environmental couplings can introduce both oscillatory effects and exponential damping to the underlying signal as discussed above.
DD can mitigate these oscillatory and exponential damping effects.
It is easiest to see these effects on the simple observables $r(t)$ used for rescaling. With an ideal quantum computer, the expected value of $r(t)$ for our studies should be $1$ regardless of the circuit depth. 
We show the effects of including DD on the rescaling circuit in Fig. \ref{fig:dynamicdecouplingeffectsrescale}.
If only Pauli or depolarizing noise channels are affecting the quantum system, then this circuit should decay exponentially with depth. 
However we observe that without DD environmental effects introduce oscillatory terms which invalidate Eq. (\ref{eq:depolarizingrescale}). 

To quantify the efficacy of the different DD sequences, we ran simulations of the rescaling circuit on \quito\,and \manila\, with CPMG, XY4, and EDD protocols. 
In order to avoid over-optimizing our choice of DD sequence, these studies used a different initial state than the one used in our final results corresponding to
 $|\psi\rangle = \frac{1}{\sqrt{2}}(|0\rangle|+\rangle|0\rangle + |1\rangle |-\rangle |1\rangle)$. The results were then fit to the ansatz

\begin{equation}
    \label{eq:rescaleansatz}
    f(x) = A e^{-B x} \cos(C x) + D,
\end{equation}
which is inspired by studies in Ref.~\cite{PhysRevLett.82.2417}. The inclusion of the oscillatory term is often seen when superposition states are prepared \cite{PhysRevApplied.18.024068}.
When comparing the fits with and without DD, we expect to see that the coefficients $B$ and $C$ should decrease when DD sequences are included into the quantum simulation. 
We find fit coefficients for the example case shown in Fig. \ref{fig:dynamicdecouplingeffectsrescale} and in this and all other cases it is indeed observed that DD lowers the fitted values of $B$ and $C$. For the example shown in Fig.~\ref{fig:dynamicdecouplingeffectsrescale}, we find that $B = 0.1356(58)$ and $C=0.3664(68)$ without DD and $B = 0.1016(18)$ and $C=-0.0919(54)$ when DD is included.
Similar trends are observed with the other simulations.

\begin{figure}[!t]
\subfigure[Rescaling circuit observable]{\includegraphics[width=\linewidth]{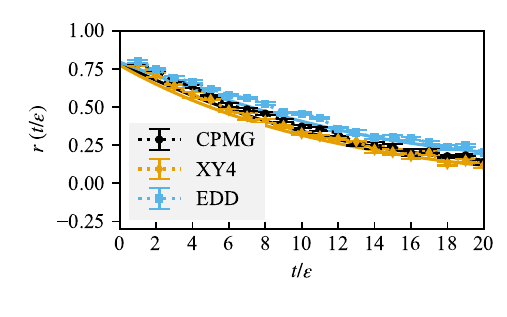}}
\subfigure[evolution circuit observable]{\includegraphics[width=\linewidth]{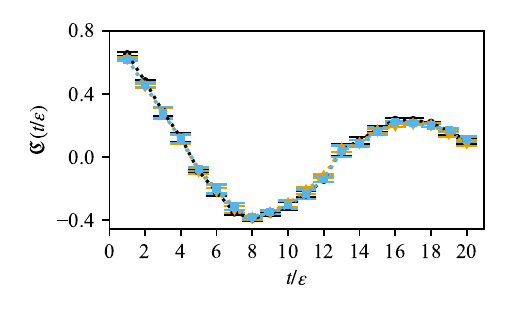}}
\caption{ Comparison of noisy estimates of the rescaling circuits and correlation functions using the three DD sequences describes in the main text for an $N_s = 2$ lattice representing [$e_0, \sigma_{0,1}, p_1$] using qubits [3, 2, 4] on \quito.}
\label{fig:ddeffectivenessq2}
\end{figure}

\begin{figure}[!t]
\subfigure[Rescaling circuit observable]{\includegraphics[width=\linewidth]{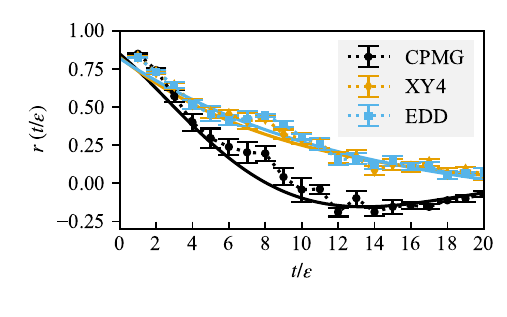}}
\subfigure[evolution circuit observable]{\includegraphics[width=\linewidth]{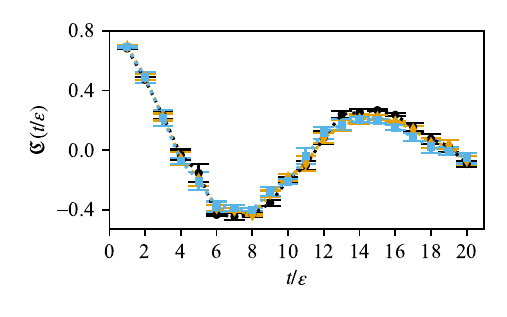}}
\caption{Comparison of noisy estimates of the rescaling circuits and correlation functions using the three DD sequences describes in the main text for an $N_s = 2$ lattice representing [$e_0, \sigma_{0,1}, p_1$] using qubits [3, 2, 4] on \manila.}
\label{fig:ddeffectivenessm2}
\end{figure}

The rescaling circuits were also fit to Eq.~(\ref{eq:rescaleansatz}) and the resulting $B$ values are shown in Figs. \ref{fig:ddeffectivenessq2}- \ref{fig:ddeffectivenessm2} and Table  \ref{tab:decoupcomp}.
The results show that XY4 and EDD generally perform better for this study than CPMG since they have a smaller value of $B$. 
There is a slight preference for EDD over XY4;
however there are physical hardware constraints that limit the ability to implement EDD due to its large number of gates. 
Thus XY4 was used for all further simulations. 

\begin{table}[!t]
\caption{Best-fit parameters and associated goodness-of-fit for fits of rescaling circuit results to Eq.~\eqref{eq:rescaleansatz} using the three dynamical decoupling sequences described in the main text. Calculations were performed using \manila~and \quito~.}
\label{tab:decoupcomp}
\begin{tabular}{cccccc}
\hline\hline
Machine & Qubits & Sequence & $B$ & $C$ & $\frac{{\chi}^2}{\text{dof}}$\\\hline
\quito & (1,0,2) & CPMG & 0.1137(72) & 0.070(43) & 0.87\\
\quito & (1,0,2) & XY4 & 0.117(72) & 0.069(58) & 2.2\\
\quito & (1,0,2) & EDD & 0.116(10) & 0.116(76) & 1.6\\
\quito & (3,2,4) & CPMG & 0.0870(37) & -0.068(18) & 0.97\\
\quito & (3,2,4) & XY4 & 0.0836(49) & 0.0006(81) & 1.4\\
\quito & (3,2,4) & EDD & 0.0725(31) & -0.075(18) & 2.2\\
\manila & (1,0,2) & CPMG & 0.1493(60) & 0.076(23) & 0.53\\
\manila & (1,0,2) & XY4 & 0.1321(46) & 0.095(14) & 0.42\\
\manila & (1,0,2) & EDD & 0.1418(54) & 0.087(15) & 0.60\\
\manila & (3,1,4) & CPMG & 0.1163(97) & 0.1939(68) & 1.4\\
\manila & (3,1,4) & XY4 & 0.1079(84) & 0.0001(31)& 1.9\\
\manila & (3,1,4) & EDD & 0.0793(60) & 0.00007(62)& 3.2\\\hline\hline
\end{tabular}
\end{table}

\section{Numerical Results}

\label{sec:numresult}

\begin{figure*}
\includegraphics[width=\linewidth]{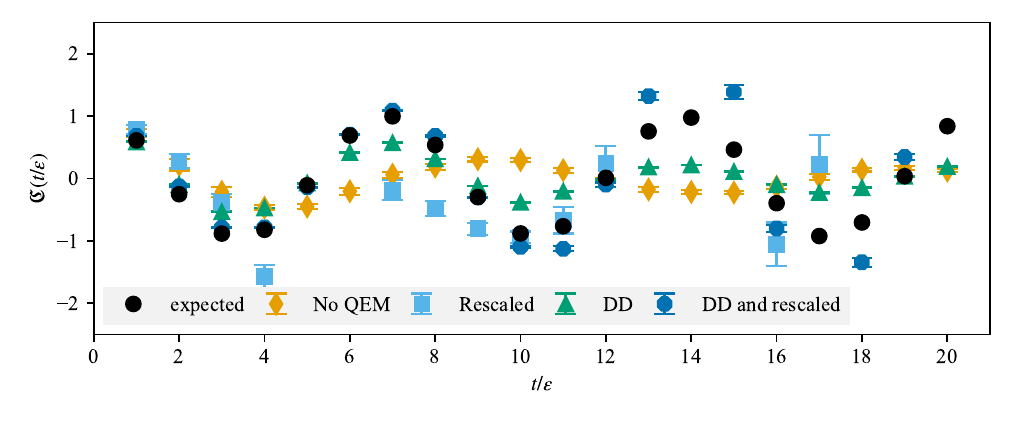}

\caption{ Effects of rescaling and/or DD on correlation-function results. The results labeled ``No QEM'' include readout mitagation and randomized compiling but neither rescaling nor DD. Errors show statistical uncertainties determined using bootstrap methods as described in the main text. The parameters for this simulation are $m_0 = 1$ and $N_s = 2$. Data points outside of the range $\pm2$ are not shown.}
\label{fig:rescalingexamplestuff}
\end{figure*}

Quantum simulation results with and without QEM are shown for one set of simulation parameters in Fig.~\ref{fig:rescalingexamplestuff}.
Exact results can be obtained classically for the lattice sizes studied in this work and are shown for comparison.
It is clear that with only randomized compiling and readout error mitigation, quantum simulation results deviate significantly from their expected values for all $t/\varepsilon > 1$. This demonstrates that randomized compiling is able to transform some but far from all systematic uncertainties into statistical uncertainties. The inclusion of DD without rescaling improves agreement between quantum simulation results and expectations for small circuit depths, but there are still statistically significant discrepancies visible for $t / \varepsilon \gtrsim 2$.  In particular the amplitude of the signal decays with increasing $t/\varepsilon$, while the expected result described a fixed-amplitude oscillation.

This decay is effectively counteracted for some values of $t/\varepsilon$ by including rescaling alone; however, at other values of $t/\varepsilon$ rescaling leads to overcorrections or drives the simulation results further away from the expected result.
The overcorrection could arise from the fact that the rescaling circuit and the time evolution circuit are not exactly the same length and the small mismatch leads to an accumulation of error as the simulation circuit depth increases. In addition, randomized compiling using only Pauli matrices does not exactly map the coherent error to a depolarizing channel.
This mapping is only true if the full Clifford group is used. 
Given this, the rescaling procedure defined in Eq. (\ref{eq:rescaleansatz}) is not exactly correct.
Rescaling will typically fail if the observable used for rescaling has any oscillatory components.

\begin{figure*}[t!]
\includegraphics[width=\linewidth]{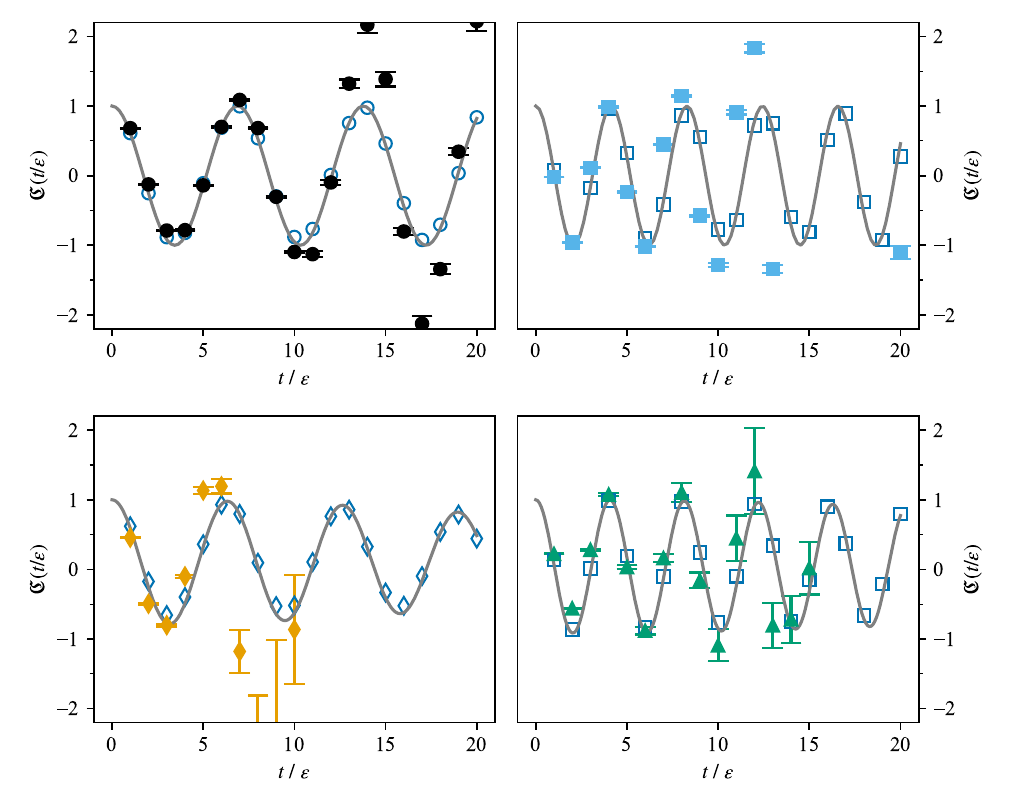}
\caption{ Fully mitigated results (including rescaling and DD) for $m_0=1$ and $N_s=2$ on \nairobi~(top left), $m_0=1$ and $N_s=4$ on \jakarta~(bottom left), $m_0=2$ and $N_s=2$ on \jakarta~(top right), and $m_0=2$ and $N_s=4$ on \nairobi~(bottom right) simulations. Points that are overly noisy, unreliable, or outside the plot range are not shown.}
\label{fig:finalized}
\end{figure*}

The simultaneous inclusion of DD and rescaling leads to significant improvements over the use of either technique alone. The quantum simulation results for $t/\varepsilon \lesssim 10$ agree with theoretical expectations within 5\% precision. Even for larger $t/\varepsilon$, results obtained with DD and rescaling are much closer to theoretical expectations than unmitigated results. These results show that the amplitude of the fully mitigated results still leaves the expected physical range for large $t/\varepsilon$ but suggest that the frequency of observed oscillations approximately matches the expected frequency.
Similar behavior can be observed in fully mitigated quantum simulation results for each set of parameters $m_0$ and $N_s$ considered here, as shown in Fig. \ref{fig:finalized}.

To quantify the accuracy of our simulation results with and without QEM, we perform fits of $\mathfrak{C}(t)$ to the spectral representation discussed in Sec.~\ref{sec:spec}.
The simplest ansatz is the single-cosine form shown in Eq.~\eqref{eq:cos}, which assumes that the initial state can be approximated as a superposition of two energy eigenstates.
The largest correction to this form arising from Trotterization is that physical times $t$ are equal to $a_t N_t$, where the renormalized Trotterization scale $a_t$ is only equal to the Trotter step size $t/\varepsilon$ for a non-interacting theory and otherwise is a function $a_t(\varepsilon,m_0,\eta,N_s)$ that must be determined for a given set of parameters by matching a dimensionful observable to a known value through scale setting.
Taking $t = a_t N_t$ in Eq.~\eqref{eq:cos} allows us to express the single-state fit ansatz as 
\begin{equation}
    \mathfrak{C}(t) \approx \cos( a_t M N_t ) e^{B t}, \label{eq:fit}
\end{equation}
where $e^{Bt}$ factor introduces a nuisance parameter $B$ in order to account for residual effects of depolarizing noise that are not completely removed by rescaling.
This shows that $a_t M$ is the dimensionless Fourier conjugate variable to the number of Trotter steps $N_t$.
Correlated $\chi^2$-minimization fits to this functional form are used to extract $a_t M$. The ansatz is then fit to all possible ranges of fit data with six or more consecutive Trotter steps. We then use Bayesian model averaging from Ref. \cite{Jay:2020jkz} to determine the time dependent mass and associated systematic and statistical uncertainties.  If these results described a physically relevant lattice gauge theory, we could match observables like $a_t M$ to their experimental values in order to determine $a_t$ and therefore make unambiguous predictions for other energies in physical units. For the 1+1 dimensional model at hand, $a_t M$ provides a proof-of-principle demonstration of the calculation of an observable that could be used for scale setting and is the final result of this work.

Results for $a_t M$ obtained from fitting our quantum simulation results at each $m_0$ and $N_s$ studied are shown in Table~\ref{tab:masses}. The theoretically expected exact results $(a_t M)_{\text{expected}}$ computed classically are shown for comparison in the same table, and $1\sigma$ agreement between quantum simulation results and these expectations is found in the $a_s m_0 = 1$ cases. However, significant discrepancies are found in the $a_s m_0 = 2$ cases. These discrepancies may arise from couplings to other excited states whose exact energies are close to the ones extracted from fits to our quantum simulations. 
It is noteworthy that applying QEM methods, and in particular the combination of rescaling and DD, is found to be necessary for achieving a good fit to Eq.~\eqref{eq:fit}.
Performing analogous fits to results without either of these techniques leads to less accurate estimates of the mass gap with larger uncertainties.

\begin{table}[!t]
    \centering
    \caption{Mass gap determined by fitting quantum simulation results to the single-state ansatz described in the main text.}
    \label{tab:masses}
    \begin{tabular}{ccccc}
    \hline\hline
         $N_s$ & $m_0$ & $\varepsilon$ & $a_t M$ & $(a_t M)_{\text{exact}}$\\
         \hline
        2 & 1.0 & 0.3 & 0.89(13) & 0.9473 \\
        4 & 1.0 & 0.3 & 1.02(18) & 0.9386 \\
        2 & 2.0 & 0.3 & 1.619(53) & 1.5204 \\
        4 & 2.0 & 0.3 & 1.591(27) & 1.5168 \\
        \hline\hline
    \end{tabular}
\end{table}

\begin{figure}[!t]
    \centering    
    \hspace{-0.8cm}\includegraphics[width=\linewidth]{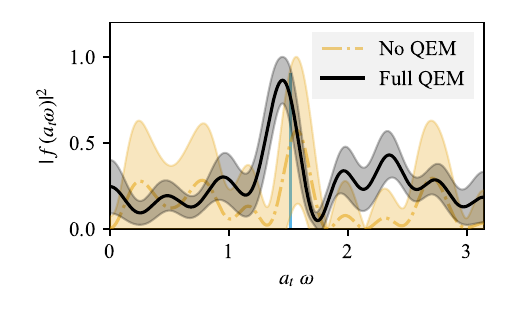}\vspace{-0.4cm}
    \caption{Fourier transform of the $m_0=2$ and $N_s=4$ correlation-function results. The yellow dash-dotted line is from the simulation only applying readout mitigation and randomized compiling, while the black solid line is from the simulation including dynamic decoupling and rescaling. The apparent oscillatory behavior of the amplitude could arise from correlations among neighboring points and ringing artifacts. The blue line indicates the exact mass gap of the Trotterized Hamiltonian.}
    \label{fig:fourierspectrum}
\end{figure}

The effect of QEM can also be studied in Fourier space by calculating the discrete Fourier transform (DFT.
\begin{equation}
  f(a_t \omega) \propto \sum_{t/\varepsilon = 0}^{N_t}\mathfrak{C}(t) \, \cos(a_t \omega t / \varepsilon).
\end{equation}
The squared magnitude $|f(a_t \omega)|^2$ for the $m_0 = 2$ and $N_s = 4$ (7 qubit lattice) correlation functions with and without DD and rescaling are shown in
Fig. \ref{fig:fourierspectrum}.
In a noiseless simulation, a clear peak around $a_t \omega \approx (a_t M)_{\text{expected}}$ should be visible that would approach a $\delta$-function as $N_t \rightarrow \infty$.
While the simulation without DD and rescaling shows an apparent peak close to this value, it is statistically not significant at 1.3 $\sigma$.
Including DD and rescaling leads to a drastic decrease in relative uncertainty in frequency space and the peak close to $(a_t M)_{\text{expected}}$ is clearly visible at 6.4$\sigma$. 
The frequencies $a_t \omega$ associated with these peaks in the Fourier spectrum correspond to energy gaps (relative to the vacuum) for states that have significant overlap with the state studied here.
As expected from the success of single-cosine fits to correlation functions in the time domain, the location of the statistically significant peak visible in mitigated results is consistent with the fitted values of $a_t M$ in Table~\ref{tab:masses} and with $(a_t M)_{\text{exact}}$.
The remaining spurious oscillations in the simulation including all error mitigation techniques could arise from correlations between different frequency DFTs and possible ringing artifacts due to the finite number of Trotter steps.

\begin{figure}
    \centering
    \includegraphics[width=\linewidth]{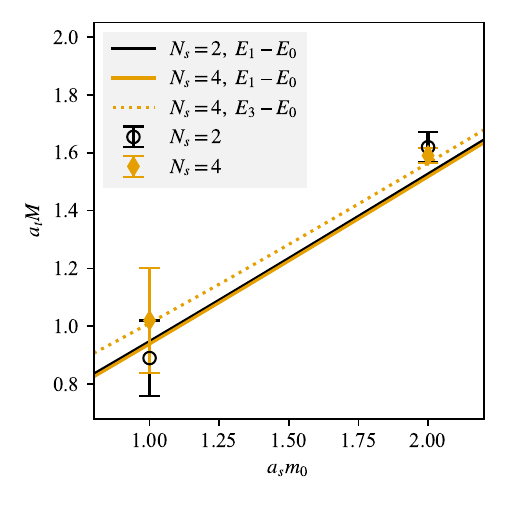}
    \caption{The mass gap in lattice units, $a_t M$, determined using quantum simulation results is shown as a function of the bare mass $a_s m_0$ as points with error bars. Solid lines indicatexact results for the Trotterized Hamiltonian}
    \label{fig:massdependence}
\end{figure}

To measure the efficacy of QEM, we use a figure of merit that quantifies how much longer a mitigated circuit can be simulated over an unmitigated one.  This definition uses the relative deviation $\delta_{\lambda}^{\mathcal{M}}(t)$ of a simulation with parameters $\lambda=\{m_0,N_s,\epsilon,\texttt{device}\}$ at a time $t/\varepsilon$ from the noiseless exact value with a mitigation strategy $\mathcal{M}$:
\begin{equation}
    \delta_{\lambda}^{\mathcal{M}}(t)=\sqrt{\frac{\sum_{t_i=0}^t\ [\mathfrak{C}_{\text{exact}}(t_i)-\mathfrak{C}_{\mathcal{M}}(t_i)]^2}{\sum_{t_i=0}^t\mathfrak{C}_{\text{exact}}(t_i)^2}},
\end{equation}
We then define $t_{\lambda}^{\mathcal{M}}(\Delta)$ as the first trotter step $t/\varepsilon$ such that $\delta_{\lambda}^{\mathcal{M}}(t)$ is larger than a threshold $\Delta$.    
An improvement factor from mitigation can then be defined as
\begin{equation}
T_\lambda(\Delta)=\frac{t_{\lambda}^{\mathcal{M}}(\Delta)}{t_{\lambda}^{\textbf{0}}(\Delta)}
\end{equation}
where $\mathcal{M}=\textbf{0}$ corresponds to the unmitigated or only randomly compiled cases that we use as baselines. This is similar to the relative error mitigation metric proposed in Refs.~\cite{2023arXiv230102690A,Cirstoiu:2022rqy}. For three of our parameter sets, we have $\mathcal{M}=\textbf{0}$ results available for comparison. We find that $1\leq T_{\lambda}(\Delta)\leq 20$ for $0\leq \Delta \leq 1$ with $T$ monotonically increasing with $\Delta$ --- here $\Delta \lesssim 0.2$ trivially shows no signs of improvement becuase $t^{\mathcal{M}}_{\lambda}(\Delta)$ is equal to 1 for all $\mathcal{M}$, while the maximum value of $T_{\lambda}$ achieved corresponds to the number of Trotter steps $N_{t}$.   Finally, to compute a single value, we average over a wide range of reasonable choices $\Delta = d/25$ with $d \in \{1,\ldots,25\}$ as
\begin{equation}
    \bar{T}=\frac{1}{N_{\Delta}N_\lambda}\sum_{\Delta,\lambda}T_{\lambda}(\Delta).
\end{equation}
This leads to $\bar{T}=5.92(12)$, which indicates that QEM enables about six times more Trotter steps to be computed with a given level of precision.

\section{\label{sec:Conclusions}Conclusions}
Quantum simulations offer promising methods for computing the non-equilibrium responses of strongly-coupled quantum systems to applied currents.
In high-energy physics, such responses can be used to predict observables ranging from transport coefficients to scattering cross sections~\cite{Bauer:2022hpo}. 
Predictions for these non-equilibrium observables require calculations of Minkowski correlation functions including the time-evolution operator $U(t)$. 
In this work, we computed a simple two-point Minkowski correlation function in $1+1d$ $\mathbb{Z}_2$ gauge theory in order to study how such observables might be determined using quantum simulations. 
After combining several quantum error mitigation strategies, we obtained an accurate description of the time dependence of the correlation function and thereby the energy of the lowest-lying spin-1 state.
These QEM strategies were essential for extracting accurate predictions from the NISQ-era quantum hardware used here. 

During our studies, we found interesting synergies between various QEM techniques.
Randomized compiling allows some hardware-level stochastic systematic errors to be converted to quantifiable statistical uncertainties, but quantum simulation results using only this technique still differ significantly from exact results.
Rescaling can be used to remove exponential signal decay with respect to circuit depth but requires a noise model that correctly describes the remaining systematics.
We find that a simple noise model including only a depolarizing channel does not adequately describe our simulation results when only randomized compiling and readout error correction are employed. However, this simple noise model satisfactorily describes our results when dynamic decoupling is included.
Leveraging all of these error mitigation strategies together,
we extend the range of simulation times over which accurate results are achieved by a factor of 6 in comparison to results using only randomized compiling and readout error correction.

Extending these simulations to realistic LGTs in four dimensions will require further theoretical and algorithmic developments.
The simplicity of $1+1d$ $\mathbb{Z}_2$ gauge theory allowed us to construct initial states that predominantly overlap with a small number of energy eigenstates.
Constructing such initial states will be increasingly difficult to achieve for more complex theories and in higher dimensions, and a variety of methods for efficient initial-state preparation are under active investigation~\cite{Kuhn:2014rha,Chakraborty:2020uhf,Kokail:2018eiw,Xie:2022jgj,Yamamoto:2021vxp,Davoudi:2022uzo,Avkhadiev:2019niu,Gustafson:2022hjf,peruzzo2014variational,Abrams:1998pd,PhysRevLett.117.010503,farhi2000quantum,Farhi472,Kaplan:2017ccd, 2010PhRvL.105q0405B,Jordan:2011ne,Jordan:2011ci,Garcia-Alvarez:2014uda,Jordan:2014tma,Moosavian:2017tkv,Lamm:2018siq,Gustafson:2019mpk,Gustafson:2019vsd,Harmalkar:2020mpd,Gustafson:2020yfe,Jordan:2017lea,Klco:2019xro,PhysRevLett.108.080402,brandao2019finite,Clemente:2020lpr,motta2020determining,deJong:2021wsd,Gustafson:2023ayr}

Trotterization errors were not found to be numerically significant for the range of times used in this analysis, but these issues are also likely to be more challenging in more complex simulations.
One promising avenue for addressing these errors is the construction of improved Hamiltonians~\cite{Carena:2022kpg,Gustafson:2022hjf} along the lines of the Symanzik improvement program used in classical LGT simulations~\cite{Symanzik:1983dc}.
Finally, it is anticipated that the efficacy of particular error mitigation strategies will change dramatically for larger gauge groups and nonabelian gauge theories where the gauge and fermion registers themselves require multiple qubits. For these cases, additional QEM methods such as those discussed in Refs.~\cite{Gustafson:2023swx,Lamm:2020jwv} may be important to include. 
Although NISQ-era quantum computers are not large enough for phenomenologically relevant lattice-QCD simulations, it is crucial to develop QEM strategies and identify best practices for Hamiltonian simulations of lattice gauge theories so that the community is ready to efficiently and effectively utilize large-scale  quantum computers when they become available.

\begin{acknowledgments}
The authors would like to thank Elias Bernreuther, Arnuad Carignan-Dugas, Pedro Machado, and Tanner Trickle.  This work is supported by the Department of Energy through the Fermilab QuantiSED program in the area of ``Intersections of QIS and Theoretical Particle Physics" and National Quantum Information Science Research Centers, Superconducting Quantum Materials and Systems Center (SQMS) under the contract No. DE-AC02-07CH11359.  
Fermilab is operated by Fermi Research Alliance, LLC under contract number DE-AC02-07CH11359 with the United States Department of Energy. F.H. acknowledges support by the Alexander von Humboldt foundation. E.G. was supported by the NASA Academic Mission Services, Contract No. NNA16BD14C. 
We acknowledge use of the IBM Q for this work. The views expressed are those of the authors and do not reflect the official policy or position of IBM or the IBM Q team. 
\end{acknowledgments}

\bibliography{refs3}

\end{document}